\documentclass[sigconf]{acmart}
\AtBeginDocument{%
  }

\usepackage{algorithm}
\usepackage{algpseudocode}
\usepackage{tikz}
\usetikzlibrary{positioning,fit,backgrounds,arrows.meta,calc,matrix}
\usepackage{fontawesome5}
\usepackage{subcaption}
\setcopyright{none}
\renewcommand\footnotetextcopyrightpermission[1]{}
\begin{document}

\title[Fleet-Scale Agentic Feature Optimization]{Cost-Aware Agentic Feature Optimization System for Fleet-Scale Ranking Models}

\author{Chang Yang}
\orcid{0000-0001-6759-6847}
\affiliation{%
  \institution{Meta, Inc.}
  \city{Sunnyvale}
  \country{USA}
}
\email{changyang@meta.com}

\author{Yao Xuan}
\affiliation{%
  \institution{Meta, Inc.}
  \city{Sunnyvale}
  \country{USA}
}
\email{yxuan@meta.com}

\author{Wendan Yan}
\affiliation{%
  \institution{Meta, Inc.}
  \city{Bellevue}
  \country{USA}
}
\email{emilywendanyan@meta.com}

\author{Nhat Le}
\affiliation{%
  \institution{Meta, Inc.}
  \city{Sunnyvale}
  \country{USA}
}
\email{nhatle@meta.com}

\author{Jiang Liu}
\affiliation{%
  \institution{Meta, Inc.}
  \city{Sunnyvale}
  \country{USA}
}
\email{jiangliu@meta.com}

\author{Longhao Jin}
\affiliation{%
  \institution{Meta, Inc.}
  \city{Sunnyvale}
  \country{USA}
}
\email{ronjin@meta.com}

\author{Peggy Yao}
\affiliation{%
  \institution{Meta, Inc.}
  \city{Sunnyvale}
  \country{USA}
}
\email{peggy@meta.com}

\author{Alireza Vahdatpour}
\affiliation{%
  \institution{Meta, Inc.}
  \city{Bellevue}
  \country{USA}
}
\email{alirezav@meta.com}

\author{Djordje Gligorijevic}
\affiliation{%
  \institution{Meta, Inc.}
  \city{Sunnyvale}
  \country{USA}
}
\email{gligorijevic@meta.com}

\begin{abstract}
Large Web platforms operate large fleets of ranking models, each selecting features from a shared pool of tens of thousands. We present an end-to-end agentic system that optimizes features across the fleet under resource constraints shared across models and those set per model, such as feature pipeline cost and serving throughput (QPS), respectively. In this system, LLM agents propose hypotheses, and deterministic tools execute experiments and verify them. For each model, an LLM generates and shortlists cross features from a combinatorial candidate space, and all features are then ranked by several importance methods. Feature blending mixes these rankings into a single selection, using Pareto fronts across tasks for multi-task models. An LLM Researcher agent then tunes the feature selection in 10--25 asynchronous trials under resource caps, and a trimming step keeps the pipeline cost within budget. Across production models, the system improves normalized entropy by $0.31\%$ on average within each model's budget, and the trimming step cuts pipeline cost by $2.3\%$ with no regression beyond measurement noise. The optimized models were launched, and in online A/B tests they improved a primary online metric by $0.20\%$.
\end{abstract}



\maketitle
\pagestyle{plain}

\section{Introduction}
Online platforms often run large numbers of ranking models catering to different surfaces and intents, while each model selects from an ever-growing pool of features. 

As every feature performs differently across models and has its own serving and infrastructure cost, feature selection is an integral component of model performance and prediction relevancy. Due to these constraints, feature selection is a cost-sensitive trade between prediction accuracy and test-time feature cost~\cite{xu2012greedy} 
that so far required manual reasoning and actions by ML practitioners. 
Recently, LLMs have been used to select~\cite{jeong2025llmselect,zhou2026mofa} 
and engineer features~\cite{hollmann2023caafe}, while LLM agents have shown increasing capabilities of running ML experimentation end to end~\cite{gao2026agentic,wang2026selfevolving}. These works study a single model in isolation. We design a scalable agentic system that generates, optimizes and selects features for every model without per-model engineering, under two kinds of budgets: per-model serving-cost caps (throughput, latency, memory) and feature pipeline budgets (processing, data, logging) shared across models. Two fleet-level mechanisms set it apart from per-model tools: a single retrieval LLM fine-tuned on feature outcomes from many models, so feature creation improves as more models are optimized; and fleet-aware pipeline-cost trimming, which optimizes
each model's features against the rest of the fleet: a feature another model under the same budget already uses adds no pipeline cost.

\begin{figure*}[t]
\centering
\begin{subfigure}[t]{0.735\textwidth}
  \centering
  \resizebox{\linewidth}{!}{\providecommand{\llmicon}{\faRobot}
\providecommand{\toolicon}{\faCog}
\providecommand{\dataicon}{\faDatabase}
\begin{tikzpicture}[
  x=1mm, y=1mm,
  font=\sffamily\footnotesize,
  box/.style={rectangle, rounded corners=3pt, align=center, minimum height=11mm,
              text width=18mm, inner sep=3pt, line width=0.6pt},
  llm/.style={box, fill=orange!15, draw=orange!75!black},
  tool/.style={box, fill=blue!10, draw=blue!60!black},
  data/.style={box, fill=green!12, draw=green!45!black},
  term/.style={box, text width=14mm, rounded corners=8pt, fill=gray!12, draw=gray!60!black},
  arr/.style={-{Stealth[length=2mm]}, line width=0.6pt, draw=black!70},
  panel/.style={rounded corners=6pt, fill=black!3, draw=black!40, dashed, inner sep=4mm},
  legend/.style={box, minimum height=5mm, text width=, font=\sffamily\scriptsize, inner sep=2pt},
]
\node[term] (inc) at (0,0) {Incumbent\\model};
\node[llm, text width=20mm] (create) at (28,9) {\llmicon\ LLM feature\\creation (\S\ref{sec:creation})};
\node[data] (pool) at (28,-9) {\dataicon\ Feature\\pool $\mathcal{F}$};
\node[tool, text width=22mm] (raf) at (62,-10) {\toolicon\ Feature ranking (\S\ref{sec:raf})\\{\scriptsize permutation, Shapley, Kernel SHAP, segment-aware}};
\node[tool, text width=22mm] (mo) at (62,9) {\toolicon\ Feature blending (\S\ref{sec:mo})};

\node[anchor=south east, inner sep=1pt, font=\sffamily\scriptsize, text=blue!60!black] at (mo.north east) {see (b)};

\node[llm] (obs) at (95,9) {\llmicon\ Observe\\trial history};
\node[llm, text width=24mm] (prop) at (129,9) {\llmicon\ Propose hypothesis\\{\scriptsize greedy $\cdot$ explore $\cdot$}\\{\scriptsize evolve $\cdot$ recipe}};
\node[tool] (val) at (129,-9) {\toolicon\ Validate\\\& trim};
\node[tool] (train) at (95,-9) {\toolicon\ Train \& measure\\NE, resources;\\log to history};
\begin{scope}[on background layer]
  \node[panel, fit=(obs)(prop)(val)(train),
        label={[align=center, font=\sffamily\footnotesize]above:{Constrained search (\S\ref{sec:agentic-fs})\\
        {\scriptsize under serving-cost caps \& shared pipeline budgets}}}] (search) {};
\end{scope}

\node[term] (final) at (165,0) {Final\\candidate};

\draw[arr] (inc.east) -- ++(2,0) |- (create.west);
\draw[arr] (inc.east) -- ++(2,0) |- (pool.west);
\draw[arr] (pool.north) -- (create.south);
\draw[arr] (create.east) -- ++(3,0) |- (raf.west);
\draw[arr] (pool.east) -- (pool.east -| raf.west);
\draw[arr] (raf.north) -- (mo.south);
\draw[arr] (mo.east) -- (obs.west);
\draw[arr] (obs.east) -- (prop.west);
\draw[arr] (prop.south) -- (val.north);
\draw[arr] (val.west) -- (train.east);
\draw[arr] (train.north) -- node[right, font=\sffamily\scriptsize] {until $K$ trials} (obs.south);
\draw[arr] (final.west -| search.east) -- node[above, font=\sffamily\scriptsize] {best} (final.west);

\node[llm, legend, anchor=west] (l1) at (-7,-24) {\llmicon\ LLM agent};
\node[tool, legend, right=2mm of l1] (l2) {\toolicon\ Deterministic tool};
\node[data, legend, right=2mm of l2] {\dataicon\ Data};
\end{tikzpicture}}
  \caption{System overview.}
  \label{fig:pipeline}
\end{subfigure}\hfill
{\color{black!20}\vrule width 0.4pt}\hfill
\begin{subfigure}[t]{0.235\textwidth}
  \centering
  \resizebox{!}{39mm}{\begin{tikzpicture}[
  x=40mm, y=40mm,
  font=\sffamily\footnotesize,
  pt/.style={circle, inner sep=0pt, minimum size=2.6mm, line width=0.5pt},
  arr/.style={-{Stealth[length=2mm]}, line width=0.7pt},
]
\colorlet{fone}{blue!60!black}
\colorlet{ftwo}{blue!35}
\colorlet{fthree}{black!40}

\draw[arr] (0,0) -- (1.1,0);
\draw[arr] (0,0) -- (0,1.1);
\node[below=2mm] at (0.55,0) {Importance for task 1, $f_{i1}$};
\node[rotate=90] at (-0.07,0.55) {Importance for task 2, $f_{i2}$};

\draw[fthree, densely dotted, line width=0.9pt] (90:0.5) arc[start angle=90, end angle=0, radius=0.5];
\draw[ftwo, dashed, line width=0.9pt] (90:0.75) arc[start angle=90, end angle=0, radius=0.75];
\draw[fone, line width=1pt] (90:1) arc[start angle=90, end angle=0, radius=1];
\foreach \a in {82, 66, 51, 37, 22, 8}
  \node[pt, fill=white, draw=fthree] at (\a:0.5) {};
\foreach \a in {84, 70, 57, 45, 33, 20, 7}
  \node[pt, fill=ftwo, draw=ftwo!60!black] at (\a:0.75) {};
\foreach \a in {86, 74, 63.5, 53.13, 45, 36.87, 24, 10}
  \node[pt, fill=fone, draw=fone!60!black] at (\a:1) {};

\node[above right, font=\sffamily\scriptsize, text=fone!80!black] at (0.6,0.8) {\bfseries A};
\node[right=1mm, font=\sffamily\scriptsize, text=fone!80!black] at (0.8,0.6) {\bfseries B};
\node[below left, font=\sffamily\scriptsize\bfseries, text=ftwo!50!black] at (45:0.75) {C};
\draw[arr, black!55, dashed] (0.585,0.78) -- (0.545,0.55);

\node[font=\sffamily\scriptsize, text=fone] at (15:1.08) {$\mathcal{P}_1$};
\node[font=\sffamily\scriptsize, text=ftwo!70!black] at (13.5:0.83) {$\mathcal{P}_2$};
\node[font=\sffamily\scriptsize, text=fthree!80!black] at (15:0.58) {$\mathcal{P}_3$};
\end{tikzpicture}}
  \caption{Feature blending: illustrative Pareto fronts for two tasks.}
  \label{fig:pareto-fronts}
\end{subfigure}
\caption{(a) The constrained search loops under resource caps until the trial budget $K$ is spent. (b) Feature blending takes whole Pareto fronts, outermost first, until $N$ features are selected; A and B share a front, and A dominates C.}
\Description{Left: a left-to-right flowchart in which the incumbent model and feature pool feed LLM feature creation, feature ranking and feature blending, followed by a dashed constrained-search loop under serving-cost caps and shared pipeline budgets (observe trial history, propose hypothesis, validate and trim, train and measure NE and resources, log to history) that outputs the final candidate. Right: an illustrative scatter plot of features by importance for two tasks with three nested quarter-circle Pareto fronts; A and B lie on the outer front, and an arrow shows A dominating C.}
\label{fig:overview}
\end{figure*}
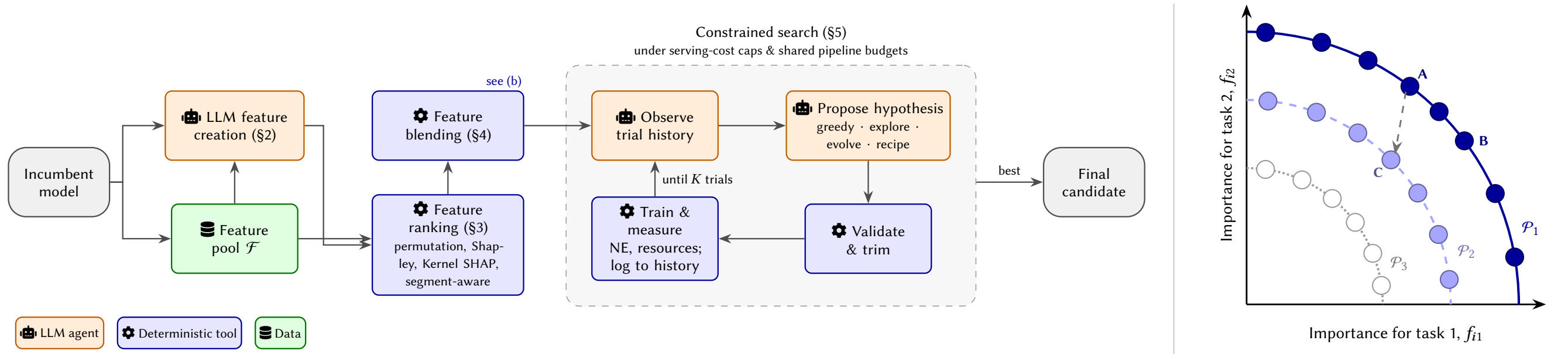

We make three contributions. (1) LLM-driven Agentic Feature Creation that uses raw signals of features to reason and generate transformed feature candidates, with one generator fine-tuned on outcomes across the fleet (Section~\ref{sec:creation}). (2) A novel feature-importance (FI) blending algorithm that combines rankings across tasks by Pareto optimality and across FI estimators, including ones scored on high-value traffic slices, by a tunable mixture (Sections~\ref{sec:raf} and~\ref{sec:mo}). (3) An agentic feature selection system that re-selects features at every refresh without manual additions or removals, balancing model quality against serving cost and the pipeline cost models share (Section~\ref{sec:agentic-fs}). Fig.~\ref{fig:pipeline} illustrates the complete system.

\section{Agentic Feature Creation}
\label{sec:creation}

\paragraph{Feature categories.}
The model input layer is built from four feature categories. \emph{Dense} features are real-valued scalars fed to multi-layer perceptron (MLP) and are the cheapest per slot. \emph{Sparse} features are lists of IDs, each mapped to a learned vector in an embedding table inside the model; these tables dominate model memory. \emph{Embedding} features are fixed-length vectors produced upstream; they need no table, but their large vectors are costly to log and store. Sparse features add table-lookup cost and embedding features add matrix-multiplication cost (a per-feature MLP projection), so which costs more serving throughput depends on the setting, such as GPU or CPU serving. \emph{Transform} features (such as the cross features) are derived inside the model from features it already receives. Feature categories differ in diminishing returns as well as cost, so improving a model usually means \emph{rebalancing} volume across feature categories rather than scaling every category up.

\paragraph{Cross-feature creation.}
Cross features (or n-gram features), the cross product of two or more ID features such as \texttt{user\_country} $\times$ \texttt{ad\_category}, often improve key metrics~\cite{He14}, but they are hand-crafted, and exhaustive search is quadratic in the pool size. Automated crossing~\cite{luo2019autocross} evaluates candidates by repeated training, which does not scale to production. We generalize crosses to \emph{feature transforms} $t = (o, U)$, which apply an operator $o \in \mathcal{O}$ (n-gram, ID matching, bucketization, etc.) to a feature subset $U \subset \mathcal{F}$ and are computed inside the model. Let $\mathcal{T}$ be the set of candidate transforms over the input feature pool $\mathcal{F}$; for bigrams alone, $|\mathcal{T}| = O(|\mathcal{O}|\,|\mathcal{F}|^2)$, which reaches tens of millions. Feature Ranking (Section~\ref{sec:raf}), a deterministic FI stage, is accurate but runs out of memory well before millions of candidates, so we use two stages. \textbf{(1)~LLM retrieval (recall).} For each feature we store static signals (name, lineage, aggregation window, coverage, an LLM-written semantic description) and model-specific signals (importance, interaction scores) in a fleet-wide Signal database. An LLM scores each candidate in a single forward pass, and the top $N_r \ll |\mathcal{T}|$ are kept. \textbf{(2)~FI re-ranking (precision).} After the shortlist of candidate features is generated, we use Feature Ranking to rank and select the final set under a budget. 
\textbf{(3)~Continual fine-tuning.} The FI scores that Feature Ranking assigns to the shortlisted cross features of many models serve as labels to continuously fine-tune a single retrieval LLM shared across the fleet, so feature creation improves as more models are optimized. Cross features further enlarge the pool beyond what any model can use, so Agentic Feature Selection (Section~\ref{sec:agentic-fs}) picks the highest-ROI subset under constraints.

\section{Feature Ranking}
\label{sec:raf}

Feature Ranking scores every candidate deterministically from a single model trained once on the full candidate pool, i.e., every feature eligible under data-governance and privacy constraints, including generated cross features
(Section~\ref{sec:creation}). Scores are normalized to sum to one, per task and in aggregate, using one of several estimators: (i) \emph{permutation importance}~\cite{Breiman01} scores a feature by how much the loss increases when its values are shuffled across an evaluation batch. It underrates correlated features,
motivating two Shapley estimators: (ii) \emph{Shapley value sampling}~\cite{Castro09}, the average loss reduction from adding a feature to those preceding it in randomly sampled orderings, and (iii) \emph{Kernel SHAP}~\cite{Lundberg17}, which estimates Shapley values as the coefficients of a weighted linear model over sampled feature coalitions. (iv) A \emph{segment-aware} variant of any above estimator trains and scores only on a small but valuable slice of traffic, such as high-value users in the top revenue buckets, to surface features that global importance can bury.

\section{Feature Blending}
\label{sec:mo}

A growing number of production ads ranking models are multi-task, predicting several outcomes (e.g., clicks and conversions)~\cite{gu2026deep}. These models make it hard to rank features globally as per-task FI scores come from losses of different magnitudes, and often different units (e.g., classification vs.\ regression heads). So a weighted sum of per-task scores depends on hand-tuned task weights that are subjective and hard to maintain across model refreshes. We therefore select features by \emph{Pareto optimality} (PO), which compares features without merging their scores.

Let a model have $J$ tasks. Feature $i$ is described by its importance vector $\mathbf{f}_i = (f_{i1}, \dots, f_{iJ})$, where $f_{ij}$ is its importance for task $j$. Feature $i$ \emph{dominates} feature $i'$ if $f_{ij} \ge f_{i'j}$ for every task $j$, with strict inequality for at least one task. The features that no other feature dominates form the first Pareto front $\mathcal{P}_1$. Setting them aside and repeating gives successive fronts $\mathcal{P}_2, \dots, \mathcal{P}_L$, stopping at the smallest $L$ with $\sum_{s=1}^{L} |\mathcal{P}_s| \ge N$, where $N$ is the feature budget. Dominance only compares scores within each task, so the procedure needs no normalization or task weights. It also keeps features that matter strongly to a single task even when they rank low elsewhere.

\emph{Feature blending} combines $M$ rankings, or \emph{sources}: the incumbent production model's ranking, whose features are proven in serving, and the ranking of any estimator of Section~\ref{sec:raf}, global or segment-specific, per task or by PO across tasks.
The $N$-feature budget is split across the $M$ sources by a mixture $\boldsymbol{\pi} = (\pi_1, \dots, \pi_M)$, $\sum_m \pi_m = 1$: in a fixed order, each source $m$ adds its highest-ranked features not yet selected until it has contributed $\mathrm{round}(\pi_m N)$, so no feature is selected twice. The mixture is tuned by the LLM Researcher agent (Section~\ref{sec:agentic-fs}).

\section{Agentic Feature Selection with Constraints}
\label{sec:agentic-fs}

In production ads ranking, a feature's value is weighed against two kinds of cost: \emph{serving cost} and \emph{feature pipeline cost} (\emph{pipeline cost}). Serving cost spans throughput, latency and memory. \emph{Throughput}, the queries per second (QPS) the hardware can serve, is tracked through its common proxy, inference compute in floating-point operations (FLOPs) per example. \emph{Latency} includes feature fetching and extraction, which takes a large share of CPU in some serving stages and slows ad delivery. \emph{Memory} is dominated by the embedding tables of sparse features: each
added sparse feature brings a table that must fit in accelerator and host memory. Pipeline cost is paid upstream for every feature a model consumes~\cite{sculley2015hidden}: \emph{processing} (the queries that generate it), \emph{data} (storage of its source data) and \emph{logging} (storage and ingestion of its logged training values). These costs are multi-dimensional, type-dependent (a variable-length sparse feature can cost far more than a dense scalar) and set-dependent rather than additive, so selecting the top features by importance up to a count cap is suboptimal under these constraints. Our agentic selection system therefore combines feature creation (Section~\ref{sec:creation}), ranking (Section~\ref{sec:raf}) and blending (Section~\ref{sec:mo}) to balance model quality against them, replacing today's practice of inheriting the incumbent's features with manual additions and removals.

\paragraph{Search space.}
Let $\mathcal{C}$ be the four feature categories of Section~\ref{sec:creation}, $\mathcal{F}_c$ the candidates of category $c \in \mathcal{C}$, $\mathcal{B}$ the banned features (privacy, compliance, coverage), and $\operatorname{Blend}_q(\cdot\,;\boldsymbol{\pi})$ the first $q$ features chosen by feature blending (Section~\ref{sec:mo}) under mixture $\boldsymbol{\pi}$. The feature set $S(\boldsymbol{\theta})$ takes $n_c$ blended features per category:
\begin{equation}
S(\boldsymbol{\theta}) = \bigcup_{c \in \mathcal{C}}
\operatorname{Blend}_{n_c}\big(\mathcal{F}_c\setminus\mathcal{B};\
\boldsymbol{\pi}\big),
\quad \boldsymbol{\theta} = (\mathbf{n}, \boldsymbol{\pi}).
\label{eq:selection}
\end{equation}
The volume vector $\mathbf{n} = (n_c)_{c \in \mathcal{C}}$ and the mixture $\boldsymbol{\pi}$ fully determine the feature set. Both are initialized from the previous model iteration but rarely re-tuned per model. We treat them as hyperparameters, so rebalancing volume across categories (Section~\ref{sec:creation}) becomes a search over $\mathbf{n}$.

\paragraph{Problem.}
With $\mathrm{NE}$ the normalized entropy (lower is better; $\Delta$NE is relative to the incumbent, and gains larger than $0.03\%$ are significant), we solve
\begin{equation}
\begin{aligned}
\boldsymbol{\theta}^\star = \arg\min_{\boldsymbol{\theta}} \quad &
\mathrm{NE}(\boldsymbol{\theta}) \\
\text{s.t.} \quad & R_k(\boldsymbol{\theta}) \le \rho_k \cdot
R_k(\boldsymbol{\theta}_0) \quad \forall k \in \mathcal{R},
\end{aligned}
\label{eq:problem}
\end{equation}
where $\boldsymbol{\theta}_0 = (\mathbf{n}_0, \boldsymbol{\pi}_0)$ is the incumbent's configuration and $\mathcal{R}$ is a set of resource metrics, with $R_k(\boldsymbol{\theta})$ the value of metric $k$ under $\boldsymbol{\theta}$ and $\rho_k$ its cap relative to the incumbent: throughput and memory, the former measured by inference FLOPs per example (e.g., $\rho_k = 1.2$) or by serving QPS (entered as $1/\mathrm{QPS}$ so that its cap is an upper bound); pipeline cost is enforced separately by trimming (below). Each evaluation takes hours to days, so a session affords only $K \approx 10$--$25$ trials, at most $W \approx 4$--$8$ of them running concurrently. As a search heuristic instead of a constraint, each category's volume is kept within an empirical range around the incumbent to avoid unpredictable extremes. 

\paragraph{Pipeline-cost trimming.} 
The caps $\rho_k$ of Eq.~\eqref{eq:problem} bound serving cost; pipeline cost is bounded by shared budgets.
A feature's pipeline cost is shared across the budgets whose models use it. A feature another model under the same budget already uses adds no pipeline cost: adding or dropping a feature changes a budget's cost only if no other model under that budget uses it. Let $\Delta_b(S)$ be the change in budget $b$'s attributed pipeline cost when the incumbent's features $I$ are replaced by $S$, and $C_b(I)$ the pipeline cost attributed to $b$ for $I$. Every trial must satisfy $\Delta_b(S) \le \tau\, C_b(I)$ for each budget $b$ the model is charged to: $\tau = 0$ holds every budget at the incumbent's cost, $\tau > 0$ allows a bounded increase and $\tau < 0$ requires a saving. Because costs are table lookups, the constraint is enforced before training (Algorithm~\ref{alg:search}, line~11) by \textsc{Trimming}, which keeps the blended selection $S(\boldsymbol{\theta})$ (Eq.~\eqref{eq:selection}) and fixes only what breaks the budget: it removes costly additions, most expensive first, and refills each slot with the best already-used feature of the same category in the blended ranking, so per-category counts, and hence the proposed configuration, are unchanged, and FLOPs essentially so.

\paragraph{Search loop.}
The search is driven by an LLM agent, the \emph{Researcher}, which runs an observe--reason--act loop~\cite{yao2023react} and uses the LLM as an optimizer %
(Algorithm~\ref{alg:search}). It \emph{observes} the trial history $\mathcal{H}$, a tuning guide $G$ with these per-category ranges, priorities and expected $\Delta$NE and FLOPs per unit, facts computed from the FI ranking, and free-text operator steering. It \emph{reasons} over them to form a hypothesis: a rebalancing of per-category volumes, a change to the mixture, or both, with a written rationale and a behavior family, namely greedy, exploration, evolutionary (combine or mutate earlier arms), or recipe, a predefined structured move (e.g., some models have roughly $10\times$ more sparse than embedding features, so a recipe removes low-ranked sparse features and adds unused, highly ranked embedding features). It \emph{acts} by submitting the proposed trial, and one long-running session keeps the full history in context as its memory. A validator rejects duplicates, out-of-range proposals and out-of-scope parameters before they consume budget.  After the initial $W$ trials, scheduling is asynchronous: a fast loop (every $\delta_{\text{fast}}$) refills each slot as soon as a trial finishes, and a slow loop (every $\delta_{\text{slow}}$) periodically measures NE and resources, early-stops trials that exceed a cap $\rho_k$, and updates the Pareto frontier. The winner is chosen by a rule fixed in advance (Algorithm~\ref{alg:search}, lines~14--15), which the LLM cannot override.

\begin{algorithm}[t]
\caption{Asynchronous feature-volume search with pipeline-cost trimming}
\label{alg:search}
\small
\begin{algorithmic}[1]
\Require $\boldsymbol{\theta}_0$, budget $K$, concurrency $W$, caps $\{\rho_k\}_{k \in \mathcal{R}}$, tolerance $\tau$, tuning guide $G$
\State $\mathcal{H} \gets \emptyset$; submit $W$ valid hypotheses from $\textsc{Researcher}(\mathcal{H}, G)$
\Repeat \Comment{every $\delta_{\text{fast}}$}
  \State add newly finished or failed trials to $\mathcal{H}$
  \If{$\delta_{\text{slow}}$ has elapsed}
    \State measure $\mathrm{NE}_h$ and $R_k(\boldsymbol{\theta}_h)$, $k \in \mathcal{R}$, of running trials
    \State early-stop any $h$ with $R_k(\boldsymbol{\theta}_h) > \rho_k R_k(\boldsymbol{\theta}_0)$ for some $k$
    \State update the NE--resource Pareto frontier
  \EndIf
  \While{$|\text{running}| < W$ \textbf{and} $|\text{submitted}| < K$}
    \State $h \gets \textsc{Researcher}(\mathcal{H}, G, \text{facts}, \text{steering})$
    \State \textbf{if} $\textsc{Validate}(h)$ \textbf{then} train $h$ on $S_h \gets \textsc{Trimming}\big(S(\boldsymbol{\theta}_h), \tau\big)$
  \EndWhile
\Until{all $K$ trials are terminal}
\State $h^\star \gets$ lowest-NE $h$ with $R_k(\boldsymbol{\theta}_h) \le \rho_k R_k(\boldsymbol{\theta}_0)\ \forall k$ and $\mathrm{NE}_h <
\mathrm{NE}(\boldsymbol{\theta}_0)$
\State \Return $(\boldsymbol{\theta}_{h^\star}, S_{h^\star})$ if $h^\star$ exists, else $(\boldsymbol{\theta}_0, I)$
\end{algorithmic}
\end{algorithm}

\begin{figure}[t]
\centering
\begin{subfigure}{\linewidth}
  \centering
  \includegraphics[width=\linewidth]{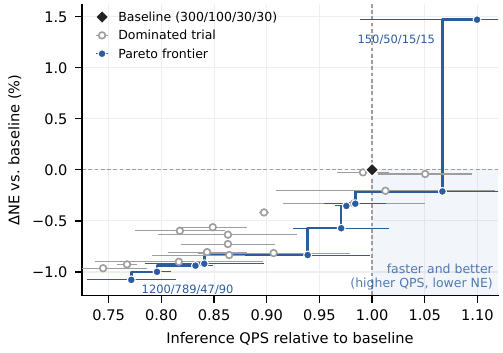}
  \caption{NE vs.\ serving QPS.}
  \label{fig:ne-qps}
\end{subfigure}

\vspace{2mm}
\begin{subfigure}{\linewidth}
  \centering
  \includegraphics[width=\linewidth]{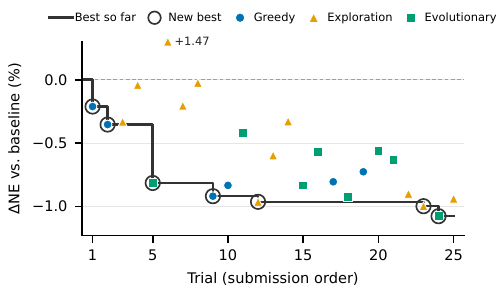}
  \caption{Search progress by trial.}
  \label{fig:best-so-far}
\end{subfigure}
\caption{Feature search (25 trials). (a) Error bars: $\pm$1 s.d.\ of QPS over 3 runs. (b) Marker shape and color give the behavior family; rings mark new bests.}
\Description{Two stacked plots. Top: scatter of NE change versus relative serving QPS for 25 trials, with error bars of one standard deviation of QPS and a step-shaped Pareto frontier. Bottom: line chart of the best NE change so far by trial index; marker shape and color give the behavior family, and rings mark new bests.}
\label{fig:search-results}
\end{figure}

\paragraph{Results.} Every trial is trained on the same data window as its baseline or control. \emph{Feature search.} These experiments tune the volumes $\mathbf{n}$ only, with $\boldsymbol{\pi}$ fixed; tuning the mixture is implemented but not yet evaluated. Table~\ref{tab:frontier} and Fig.~\ref{fig:ne-qps} show 25 trials on a baseline with 300\slash100\slash30\slash30 dense\slash sparse\slash transform\slash embedding features: the frontier runs from the fastest trial ($+1.47\%$ NE at $1.10\times$ QPS) to the most accurate ($-1.08\%$ NE at $0.77\times$ QPS), and Fig.~\ref{fig:best-so-far} shows the search keeps finding new bests late in the budget, from greedy, exploration and evolutionary trials. On a separate, production per-event conversion model (1200\slash789\slash50\slash47) under a $1.2\times$ FLOPs cap, every initial trial that only raised dense, sparse or transform volume was flat or worse, while a recipe trial that doubled embeddings and cut sparse by 14\% reduced NE by $0.17\%$ against the baseline: the gain comes from \emph{rebalancing} across categories, which inherited per-category counts never do. Across seven high-impact models, the search reduces NE by $0.31\%$ on average within each model's budget.
\emph{Pipeline-cost trimming.} Compared with an untrimmed, importance-only control on six models, trimming brings every model within budget at $\tau = 0$ and cuts attributed pipeline cost by $2.3\%$ on average (up to $6.3\%$), while keeping per-category feature counts identical, FLOPs within $0.3\%$, and main-task loss no worse than $+0.03\%$, within run-to-run noise.
\emph{Online.} 
The system has optimized models for over six months, with multiple launches; in online A/B tests, the optimized models improved a primary online metric by $0.20\%$.

\begin{table}[t]
\centering
\small
\setlength{\tabcolsep}{4pt}
\caption{Selected NE--QPS Pareto-frontier trials (full frontier in Fig.~\ref{fig:ne-qps}). Volumes are dense/sparse/transform/embedding counts; FLOPs, QPS and $\Delta$NE are relative to the baseline.}
\label{tab:frontier}
\begin{tabular}{@{}lllrrr@{}}
\toprule
Trial & Volumes & Family & FLOPs & QPS & $\Delta$NE \\
\midrule
Baseline & 300/100/30/30 & --- & 1.00$\times$ & 1.00$\times$ & --- \\
\midrule
h006 & 150/50/15/15 & explore & 0.61$\times$ & 1.10$\times$ & $+1.47\%$ \\
h001 & 600/100/30/30 & greedy & 1.07$\times$ & 1.07$\times$ & $-0.21\%$ \\
h003 & 300/100/30/60 & explore & 1.25$\times$ & 0.98$\times$ & $-0.33\%$ \\
h010 & 1200/200/60/60 & greedy & 1.82$\times$ & 0.94$\times$ & $-0.83\%$ \\
h024 & 1200/789/47/90 & evol. & 4.07$\times$ & 0.77$\times$ & $\mathbf{-1.08\%}$ \\
\bottomrule
\end{tabular}
\end{table}

\section{Conclusion}
We presented an agentic system that optimizes the features of every model it manages in an ads ranking fleet under serving and shared pipeline-cost constraints. A single LLM, fine-tuned on feature outcomes across the fleet, generates cross features; the LLM Researcher finds gains by rebalancing volume across feature categories rather than only scaling it up; and trimming keeps shared pipeline cost within budget. On production models, the system improves NE by $0.31\%$ and cuts pipeline cost by $2.3\%$; in A/B tests, it improved a primary online metric by $0.20\%$.


\end{document}